\documentclass[12pt, a4paper]{article} 

\usepackage{geometry} 
\usepackage{graphicx}
\usepackage{amsmath} 
\usepackage{amssymb} 
\usepackage{booktabs}
\usepackage{authblk}
\usepackage{hyperref} 
\usepackage[compress]{natbib}
\usepackage{setspace}
\usepackage{indentfirst}      
\usepackage{xcolor}

\title{Nonlocal correlations of fermionic entanglement in the spacetime of Einstein-Gauss-Bonnet black hole}

\author[1]{Yifei Xu}
\author[1]{Yanjun Chen}
\author[1]{Qi Xiao}
\author[1]{Xiaolong Gong\thanks{Corresponding author: xlgong@yangtzeu.edu.cn}}

\affil[1]{School of Physics and Optoelectronic Engineering, Yangtze University, Jingzhou 434023, China}

\begin{document}
\maketitle 
\vspace{-2em}
\begin{abstract}
The investigation of nonclassical correlations in curved spacetimes offers key insights into the intersection of quantum information theory and gravitational physics. This paper studies two nonlocal correlation measures, non local advantage of quantum coherence (NAQC) and Bell nonlocality (BN) in a $d$-dimensional spherically symmetric Einstein-Gauss-Bonnet (EGB) black hole spacetime. We consider two observers (Alice and Rob) initially sharing a maximally entangled Bell state: Alice freely falls into the black hole (inertial Kruskal frame), while Rob accelerates outside the horizon (non-inertial Schwarzschild-like frame). The Unruh-Hawking effect modifies Rob's field modes, requiring Bogoliubov transformations to relate the two frames. We derive the mixed bipartite density matrix for fermionic fields and analytical expressions for NAQC and BN, which depend on Hawking temperature (itself governed by $\alpha$, $d$, and $r_h$). Our results show both correlations degrade monotonically with increasing Hawking temperature, confirm the NAQC-BN hierarchical relationship persists in EGB spacetime, and highlighting the impact of high curvature corrections on quantum resources. 
\end{abstract}
	
\section{Introduction}
	\label{sec:1}
The study of nonclassical resources within the relativistic framework, including quantum coherence \cite{streltsov2017colloquium}, quantum entanglement \cite{horodecki2009quantum}, quantum steering \cite{uola2020quantum}, and Bell nonlocality \cite{brunner2014publisher}, has emerged as one of the most fascinating frontiers in quantum information theory. In contemporary research, nonlocal correlations derived from quantum coherence are regarded as essential physical resources in various quantum information processing tasks, including quantum teleportation \cite{zeilinger2000quantum}, quantum metrology \cite{giovannetti2006quantum}, quantum cryptography \cite{bennett1992quantum}, and dense coding \cite{mattle1996dense}. However, nonlocal correlations are extremely fragile and can be easily disrupted by the inevitable interaction between quantum systems and their surrounding environment \cite{xiao2024experimental}. In recent years, the nonlocal advantage of quantum coherence (NAQC) in flat spacetime has attracted widespread attention in both theoretical and experimental fields \cite{mondal2017nonlocal}. This measurement is based on the coherence complementarity relationships and establishes criteria to achieve NAQC through various coherence measures, including coherence $l_{1}$, relative entropy, and skew information \cite{datta2018sharing,mondal2017complementarity}. The theoretical significance of NAQC lies not only in revealing the key link between quantum nonlocality and the quantum speed limit but also in clarifying the hierarchical relationships among nonlocal correlations. Furthermore, its criterion can be interpreted as an entanglement witnessing \cite{hu2018hierarchy,hu2018nonlocal,ding2019experimental}, providing a new approach for experimental selection of suitable quantum resources for quantum information simulation tasks.
In recent decades, with the deepening of research at the intersection of quantum information theory and gravitational physics, the study of non-local correlations in curved spacetimes has become a frontier topic \cite{yao2025hierarchical,wang2020generation,2018AnPhy.390..334L,peres2004quantum,mann2012relativistic,harikrishnan2022accessible,du2024basis,yang2023investigating,li2024bosonic,Wu:2024yop,gross202350,zeng2024can,Xiao:2025flt,Li:2026wbt,Li:2025jlu,Li:2024pwo,Huang:2024vyc,Shang:2025ljm}. The extreme gravitational environment of black holes, characterized by strong spacetime curvature and Hawking thermal radiation, provides a unique platform to test the robustness of quantum resources and explore the compatibility between quantum mechanics and general relativity \cite{hawking1974black,hawking1975particle,schoutens1993quantum,bekenstein1980black,Liu:2024yrf,Wu:2025pxm,An:2024hdv,Hu:2021bam,Wang:2016eye,2020NatSR..1014697L,brustein2014origin,wu2024genuinely,2023EPJP..138..360G}. These investigations contribute to our understanding of several key issues in quantum field theory in curved spacetime, including quantum effects related to black hole entropy and the black hole information paradox, as well as the geometric structure of the early universe.

The idea that spacetime may have more than four dimensions has now become a standard hypothesis in high-energy physics \cite{green1987superstring}. Brane cosmology, which is compatible with string theory, argues that matter and gauge interactions may be localized on a brane embedded in a higher-dimensional space, such that the gravitational field can propagate throughout the entire spacetime. Consequently, we need to consider higher-dimensional theories of gravity. In this context, the Lovelock theory of gravity can be adopted, which features a more general form of the action incorporating higher powers of the Riemann tensor and its derivatives \cite{lovelock1971einstein}. The first two terms in the field equations correspond to the Einstein tensor and the cosmological constant, respectively, while the third term includes the square of the curvature, namely the Gauss-Bonnet tensor. At this order, the resulting field equations are referred to as the Einstein-Gauss-Bonnet gravity field equations. Many authors have obtained various solutions to these equations by assuming specific symmetries for the metric. In Ref \cite{dehghani2004asymptotically}, asymptotically AdS solution of Gauss-Bonnet gravity have been derived in the absence of a cosmological constant. At the same time, studies indicate that an accelerating expanding universe can be obtained through the modified Friedmann equations in Gauss-Bonnet gravity \cite{dehghani2004accelerated}. Therefore, the Gauss-Bonnet term appears to have an anti-gravitating effect.

Among various nonlocal correlation measures, NAQC and Bell nonlocality (BN) have attracted extensive attention due to their distinct physical connotations and practical significance: BN directly reflects the violation of local realism, while NAQC establishes a crucial link between quantum coherence and nonlocality, revealing the hierarchical structure of quantum resources. Understanding the behavior of these two measures in Einstein-Gauss-Bonnet (EGB) black hole spacetime is not only important for clarifying the influence of higher-curvature gravity on quantum correlations but also provides new insights into the quantum nature of black holes and the information paradox.

This study focuses on two types of quantum correlations, NAQC and Bell nonlocality, aiming to evaluate non-local correlation in the spacetime of an Einstein-Gauss-Bonnet black hole. We assume that two observers, Alice and Rob, initially share a two-mode maximally entangled state, where Alice freely falls into the black hole, while Rob accelerates and remains near the event horizon. We perform calculations using the composite state between them. This paper is divided into five parts. Section \ref{sec:2} briefly reviews the theory of Gauss-Bonnet gravity and analyzes the Hawking–Unruh effect for black hole solutions within this theoretical framework. Section \ref{sec:3} introduces the detailed calculation methods for NAQC and BN. Section \ref{sec:4} presents the calculations of NAQC and BN for fermionic entanglement, along with plotted figures and corresponding analysis. Section \ref{sec:5} provides concluding remarks.

 {Although a large number of prior works have investigated NAQC and Bell nonlocality in Schwarzschild, Reissner-Nordström and dilaton black hole backgrounds with identical Bogoliubov transformation frameworks, those studies only cover 4-dimensional spacetime without higher-curvature corrections. The present work focuses on $d$-dimensional Einstein-Gauss-Bonnet gravity with fermionic fields, and we reveal three qualitatively unique physical consequences absent in standard general relativity: the divergent horizon-radius dependence of quantum correlations between $d=5$ and $d\geq6$, decoherence suppression originating from the antigravitational Gauss-Bonnet coupling $\alpha$, and competitive interplay between spacetime dimensionality and high-curvature modification. These new qualitative results constitute the core novelty of our work, rather than merely substituting the metric function and Hawking temperature expression.}

\section{Quantum field near the Einstein-Gauss-Bonnet black hole}
	\label{sec:2}
	 The field equation of Einstein-Gauss-Bonnet (EGB) gravity is
	 \begin{equation}\label{1}
	 	G^{(E)}_{\mu\nu}+\Lambda g_{\mu\nu}+\alpha G^{(GB)}_{\mu\nu}=T_{\mu\nu},
	 \end{equation} 
	  where $G^{(E)}_{\mu\nu}$ is the Einstein tensor and $G^{(GB)}_{\mu\nu}$ is the Gauss-Bonnet tensor, which is defined as
	 \begin{equation}\label{2}
	 	\begin{aligned}
	 		G_{\mu \nu}^{(\mathrm{GB})} = & 2\left( R_{\mu \sigma \kappa \tau} R_{\nu}^{\ \sigma \kappa \tau} - 2 R_{\mu \rho \nu \sigma} R^{\rho \sigma} - 2 R_{\mu \sigma} R_{\ \nu}^{\sigma} + R R_{\mu \nu} \right) \\
	 		& - \frac{1}{2} \left( R_{\mu \nu \sigma \kappa} R^{\mu \nu \sigma \kappa} - 4 R_{\mu \nu} R^{\mu \nu} + R^{2} \right) g_{\mu \nu}.
	 	\end{aligned}
	 \end{equation}
	 $\alpha$ is the Gauss-Bonnet coupling constant and we take $\alpha>0$. Now we consider a d-dimensional ($d\geq5$) static spherically symmetric spacetime with the metric
	\begin{equation}\label{3}
		\mathrm{d}s^2 = -f(r)\mathrm{d}t^2 +\frac{1}{f(r)}\mathrm{d}r^2 +r^2\mathrm{d}\Omega_{d - 2}^2,
	\end{equation}
	 where $f(r)$ is an unknown metric function and $r^2\mathrm{d}\Omega_{d - 2}^2$ is the metric of a ($d-2$)-dimensional subspace. It can be proved that this metric describes a black hole solution of the field equation \eqref{1} with $\Lambda=0$, provided that
	\begin{equation}\label{4}
		f(r) = k + \frac{r^2}{2(d - 3)(d - 4)\alpha}
		\left(1 \pm \sqrt{1 + \frac{4(d - 3)(d - 4)\alpha m}{r^{(d - 1)}}}\right),
	\end{equation}
	 where $k$ is the curvature of the ($d-2$)-dimensional subspace. For the special case of $d=5$, the function takes a particular form 
	\begin{equation}\label{5}
		f(r) = k + \frac{r^2}{4\alpha} \pm \sqrt{\frac{r^4}{16\alpha^2} + \left(|k| + \frac{m}{2\alpha}\right)},
	\end{equation}
	 which has a geometrical mass $m+2\alpha|k|$. We are interested in asymptotically flat spacetime, so we set $k=1$ and choose minus sign in Eqs. \eqref{4} and \eqref{5}. In the limit $\alpha\to0$, the metric become a $d$-dimension Schwarzschild solution, as a requirement of Lovelock gravitation theory. Evidently, the radius of the horizon denoted by $r_{h}$ is determined by the largest root of $f(r)=0$. For example, $r_{h}=\sqrt{m}$ in the special case of $d=5$.
	\begin{figure}[htbp]
		\centering
		\includegraphics[width=0.8\textwidth]{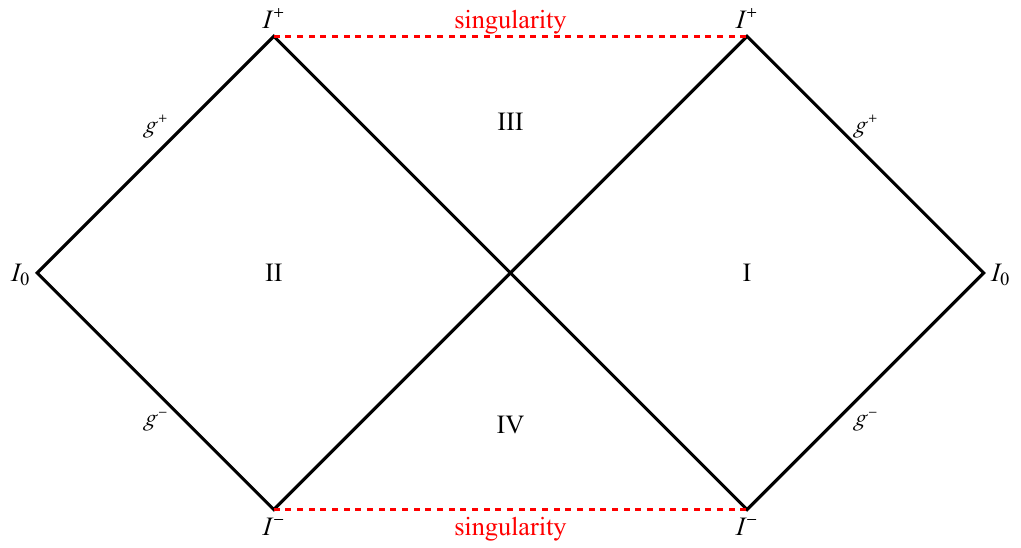} 
		\caption{Penrose diagram of metric in Eq. \eqref{3}} 
		\label{fig:penrose}
	\end{figure}
	
	 We consider the Penrose diagram of the metric in Eq. \eqref{3}, and invoke the Kruskal coordinates by using the appropriate coordinate transformations. Therefore, we write the metric in ($u,v$) coordinates as
	\begin{equation}\label{6}
		\begin{array}{l}
			ds^2 = -f(r)dudv + r^2 d\Omega_{d - 2}^2, \\
			u = t - r_*,\quad v = t + r_*,
		\end{array}
	\end{equation}
	 where $r_{*}=\int dr/f(r)$ is the Regge-Wheeler tortoise coordinate. $r_{*}$ has a logarithmic singularity at the horizon. this behavior of the tortoise coordinate near the horizon is generally establish for the general relativity(GR) theory  \cite{leonhardt2002laboratory} and for the Schwarzschild spacetime in this theory, $r_{*}$ around the radius of horizon can be expanded. For the spacetime of Gauss-Bonnet black hole, $r_{*}$ near the horizon can be expanded as
	\begin{equation}\label{7}
		r_{*}\approx \Gamma \ln (r - r_{h}) + \mathcal{G}(r - r_{h}),
	\end{equation}
	 where $\mathcal{G}(r - r_{h})$ is a nonsingular function at $r_{*}$. Comparing $\Gamma$ with the analogous coefficient in the Schwarzschild spacetime, we infer that the $\Gamma$ is the inverse of Hawking temperature of Gauss-Bonnet black hole,i.e., $\Gamma^{-1}=T$. $T$ can be defined geometrically in terms of the surface gravity of black hole as
	\begin{equation}\label{8}
		T = 2\pi \left(-\frac{1}{2}\nabla_{\mu}\xi_{\nu}\nabla^{\mu}\xi^{\nu}\right)^{-\frac{1}{2}},
	\end{equation}
	 which as applied for the spacetime \eqref{3}, lead to
	\begin{equation}\label{9}
		T = \frac{1}{4\pi} \left. \frac{df(r)}{dr} \right|_{r_h}.
	\end{equation}
	 This as evaluated for Eqs. \eqref{4} and \eqref{5}, lead to
	\begin{equation}\label{10}
		T = \frac{1}{4\pi} \frac{\alpha d^3 - 12\alpha d^2 + 47\alpha d + dr_h^2 - 60\alpha - 3r_h^2}{r_h(2\alpha d^2 - 14\alpha d + 24\alpha + r_h^2)},
	\end{equation}
	for the $d$-dimensional spacetime.  {We derive Eq. \eqref{10} strictly by substituting $f(r)$ and its radial derivative at $r=r_{h}$ into the surface-gravity temperature formula Eq. \eqref{9}. Under geometric units $G=\hbar=c=1$, $\alpha$ has dimension of length-squared, $r_{h}$ carries dimension of length, and the spacetime dimension $d$ is dimensionless, so the final expression of $T$ satisfies consistent dimensional analysis with inverse length units.} In the special case of $d=5$,
	\begin{equation}\label{11}
		T = \frac{1}{2\pi} \frac{r_h}{(4\alpha + r_h^2)}.
	\end{equation}
     {The positive GB coupling term acts as an antigravitational correction: it locally weakens spacetime curvature near the horizon, reduces surface gravity and lowers the Hawking temperature $T$. A smaller $T$ generates fewer thermal particle excitations from the Unruh-Hawking effect, hence mitigating thermal decoherence of the shared fermionic entanglement.}
	
	 Kruskal coordinates are defined as
	\begin{equation}\label{12}
		U \propto \pm e^{-uT},\qquad V \propto \mp e^{vT},
	\end{equation}
	 which are used for an analytical extension of the metric. The upper (lower) sign refers to the Region I (II) in Figure \ref{fig:penrose} which represents the Penrose diagram of metric \eqref{3}. Notice that the Regions I and II are causally disconnected.  {The topological structure of the Penrose diagram in Figure \ref{fig:penrose} coincides with that of standard Schwarzschild black holes. The Gauss-Bonnet coupling $\alpha$ and dimension $d$ only quantitatively modify the horizon radius and surface gravity, without changing the global causal separation between Region I and Region II. Even in the large-$\alpha$ limit, no extra horizons or naked singularities emerge, and the tortoise coordinate still holds logarithmic divergence at the event horizon.}
	
	 We assume that there are two observers, Alice (denotes as A) and Rob (denotes as R), who are going to communicate through a quantum information protocol in the spacetime of EGB black hole. They share a Bell state, which is composed of two modes of a free quantum field, say ground state and the first exited state of a scalar or a spinor field. A and R meet each other and share the prescribed quantum state at the asymptotic region of the black hole. Then, A stays on a timelike geodesic of the black hole, and consequently is freely falling as an internal observer. But R approaches the horizon and barely accelerates to avoid falling into the black hole. In order to express the entanglement between A and R, we need to construct the quantum field modes as seen by A and R.
	
	 We consider a massless scalar field $\phi$ that satisfies the Klein-Gordon equation
	\begin{equation}\label{13}
		\frac{1}{\sqrt{-g}}\partial_{\mu}(\sqrt{-g}g^{\mu \nu}\partial_{\nu}\phi) = 0,
	\end{equation}
	 where $g$ is the metric determinant. Regarding the spherical symmetry of the metric, $\phi$ can be separated as
	\begin{equation}\label{14}
		\phi (t,r,\Omega) = e^{-i\omega t}\frac{R_{\omega l}(r)}{r^{\frac{d - 2}{2}}} Y_{lm}(\Omega),
	\end{equation}
	 where $Y_{lm}(\Omega)$ is the ($d-2$)-dimensional spherical harmonic  functions and $R_{\omega l}(r)$ satisfies
	\begin{equation}\label{15}
		\frac{\partial^2 R_{\omega l}}{\partial r_*^2} + \omega^2 R_{\omega l} - f(r) \left( \frac{(d - 2)^2}{4r^2} f(r) + \frac{d - 2}{2r} \frac{df}{dr} + \frac{l(l + d)}{r^2} \right) R_{\omega l} = 0.
	\end{equation}

 {We fix $\omega=0.01$ for all numerical simulations and adopt the single-mode approximation. In this low-frequency limit, the de Broglie wavelength of fermionic modes is far larger than the curvature scale near the event horizon, so intermode mixing induced by the effective potential is exponentially suppressed. The effective potential in Eq. \eqref{15} mainly modifies high-frequency wavefunctions, while low-energy modes are barely disturbed by variations of $d$ and $\alpha$. We have tested multiple small values $\omega=0.001$, 0.01 and 0.1, and all parameter curves share identical qualitative trends with only minor quantitative shifts, which verifies the robustness of the single-mode approximation within our low-energy setup.}
    
	 The observer A who is freely falling into the black hole, sees nothing special at the horizon, so he has access to the entire of the spacetime. But as Figure. \ref{fig:penrose} shows, for the accelerated observer R, there are two causally disconnected regions of spacetime denoted by I and II. Since the future directed timelike killing vector corresponding to the region II is directed in the opposite direction of that of the region I, that is $[\partial_{t}]_{I}=[-\partial_{t}]_{II}=[\partial_{-t}]_{II}$, then the positive frequency solutions related to the regions I and II differ up to minus sign in $t$. Indeed the positive frequency solutions of Eq. \eqref{15} are obtained as
	\begin{equation}\label{16}
		\begin{aligned}
			\phi_{\mathrm{I},k} &\sim e^{ikr_* - i\omega t} \equiv e^{i\omega u}, \\
			\phi_{\mathrm{II},k} &\sim e^{ikr_* + i\omega t} \equiv e^{-i\omega u}.
		\end{aligned}
	\end{equation}
	 In this way any quantum field can be quantized as
	\begin{equation}\label{17}
		\Phi = \sum_{l,m}\int d\omega \left[ \left( a_{\mathrm{I}} \phi_{\mathrm{I},k} + a_{\mathrm{II}} \phi_{\mathrm{II},k} \right) + \text{H.C.} \right],
	\end{equation}
	 where $a_{\mathrm{I},k}$($a_{\mathrm{II},k}$) and$a_{\mathrm{I},k}^{\dagger}$($a_{\mathrm{II},k}^{\dagger}$) are the annihilation  and creation operators for the mode $k$ in the region I(II). These operators are called the Schwarzschild operators and satisfy
	\begin{equation}\label{18}
		\begin{aligned}
			a_{\mathrm{I},k}|0\rangle_{\mathrm{I},k}\otimes |n\rangle_{\mathrm{II}} &= a_{\mathrm{II},k}|n\rangle_{\mathrm{I}}\otimes |0\rangle_{\mathrm{II},k} = 0,\\
			a_{\mathrm{I},k}^{\dagger}|0\rangle_{\mathrm{I},k}\otimes |n\rangle_{\mathrm{II}} &= |1\rangle_{\mathrm{I},k}\otimes |n\rangle_{\mathrm{II}},\\
			a_{\mathrm{II},k}^{\dagger}|n\rangle_{\mathrm{I}}\otimes |0\rangle_{\mathrm{II}} &= |n\rangle_{\mathrm{I}}\otimes |1\rangle_{\mathrm{II},k}.
		\end{aligned}
	\end{equation}
	
	 Since the solution \eqref{16} cannot be analytically continued from the Region I to the Region II, we have to express them in the Kruskal coordinates. Regarding Eqs. \eqref{12} and \eqref{16}, we can show that the analytical solutions in the whole of the spacetime can be written as  \cite{birrell1982quantum}
	\begin{equation}\label{19}
		\begin{aligned}
			\phi_{K,k}^{+} &= e^{\frac{\pi n}{2T}}\phi_{\mathrm{I},k} + e^{-\frac{\pi n}{2T}}\phi_{\mathrm{II},-k}^{+},\\
			\phi_{K,k}^{-} &= e^{-\frac{\pi n}{2T}}\phi_{\mathrm{I},-k}^{+} + e^{\frac{\pi n}{2T}}\phi_{\mathrm{II},k},
		\end{aligned}
	\end{equation}
	 which correspond to positive and negative frequencies with respect to the Killing vector $\partial_{U}$. For a fermionic field, instead of Eq. \eqref{17}, one can now expand $\Phi$ in terms of $\phi_{K,k}^{+}$ and $\phi_{K,k}^{-}$ as
	\begin{equation}\label{20}
		\Phi = \sum_{l,m}\int d\omega \left[ \left( b_{K,k}^{+} \phi_{K,k}^{+} + b_{K,k}^{-} \phi_{K,k}^{-} \right) + \text{H.C.} \right],
	\end{equation}
	 where
	\begin{equation}\label{21}
		\begin{aligned}
			b_{K,k}^{+} &= (\cos \zeta) a_{\mathrm{I},k} - (\sin \zeta) a_{\mathrm{II},-k}^{\dagger},\\
			b_{K,-k}^{-\dagger} &= (\sin \zeta) a_{\mathrm{I},k} + (\cos \zeta) a_{\mathrm{II},-k}^{\dagger},
		\end{aligned}
	\end{equation}
	 where $\tan \zeta = e^{-\pi \omega / T}$. The ground state and the first excited states of the field in Kruskal and Schwarzschild coordinates are related by
	\begin{equation}\label{22}
		\begin{aligned}
			|0\rangle_K &= (\cos \zeta) |0\rangle_{\mathrm{II}} \otimes |0\rangle_{\mathrm{I}} + (\sin \zeta) |1\rangle_{\mathrm{II}} \otimes |1\rangle_{\mathrm{I}}, \\
			|1\rangle_K &= |1\rangle_{\mathrm{I}} |0\rangle_{\mathrm{II}}.
		\end{aligned}
	\end{equation}

     {The scalar-field mode transformation in Eq. \eqref{19} is determined purely by spacetime geometry and Kruskal-Schwarzschild coordinate mapping. For $d$-dimensional Dirac spinor fields, the coordinate transformation law of positive/negative frequency modes remains identical to scalar fields, despite different anticommutation relations for fermionic annihilation and creation operators. The parameter $tan\zeta=e^{-\pi\omega/T}$ originates from thermal periodicity and does not rely on spin degrees of freedom, consistent with the curved-space spinor quantization framework in \cite{birrell1982quantum}.}

	 {We omit greybody factors throughout this work for clear physical analysis. Greybody factors describe the tunneling probability of Hawking particles through the radial effective potential barrier, which would introduce complex frequency-dependent integrals and obscure the pure geometric effects brought by $\alpha$, $d$ and $r_{h}$. Our core target is to qualitatively reveal how higher-curvature corrections and extra dimensions modulate nonlocal correlations via Hawking temperature $T$. The full incorporation of greybody factors and quantitative spectral corrections will be considered in our follow-up research.}
\section{Non-local correlation measures}
\label{sec:3}
Quantifying non-local correlations is essential for analyzing the degradation and evolution of quantum resources under the influence of gravitational effects.
In this work, we adopt two representative quantifiers, namely the NAQC and BN,
to characterize the non-classical correlations of bipartite fermionic systems in the Einstein-Gauss-Bonnet black hole spacetime.
This section provides a detailed introduction to the definition, physical meaning, and calculation methods of these two measures.

\subsection{Nonlocal Advantage of Quantum Coherence}
The NAQC reveals the non-local advantage enabled by quantum coherence in bipartite systems,
serving as a critical indicator to distinguish non-local coherence from local classical coherence.
Based on the complementarity relation of quantum coherence, NAQC is defined as \cite{mondal2017nonlocal}
\begin{equation}\label{23}
N^{l_1}(\rho_{AB}) = \frac{1}{2}\sum_{i,j,b} p\left(\rho_{\Pi_{j\neq i}^{b}}\right) C_{l_1}^{\sigma_i}\left(\rho_{B|\Pi_{j\neq i}^{b}}\right) > C_{\text{max}},
\end{equation}
where:
\begin{itemize}
\item[(1)] $\rho_{AB}$ denotes the bipartite quantum state shared by two observers;
\item[(2)] The summation runs over $i,j \in \{x,y,z\}$ ($j \neq i$, corresponding to Pauli operators) and $b \in \{0,1\}$ (measurement outcomes);
\item[(3)] $p\left(\rho_{\Pi_{j\neq i}^{b}}\right)$ represents the probability of obtaining the measurement result $b$ when performing the $\sigma_j$ measurement;
\item[(4)] $C_{l_1}^{\sigma_i}\left(\rho_{B|\Pi_{j\neq i}^{b}}\right)$ is the $l_1$-norm coherence of the conditional state of Bob in the $\sigma_i$ basis;
\item[(5)] $C_{\text{max}} = \sqrt{6}$ is the maximum classical bound of the total coherence in three mutually unbiased bases.
\end{itemize}

For an arbitrary two-qubit state $\rho_{AB}$, the local projection measurement performed by the observer is
\[
\Pi_i^b = \frac{\mathbb{I} + (-1)^b \sigma_i}{2}, \quad i = x,y,z,
\]
with the corresponding measurement probability
\[
p_{\Pi_i^b} = \text{Tr}\left[(\Pi_i^b \otimes \mathbb{I})\rho_{AB}\right].
\]
After the local measurement, the conditional quantum state of the other observer is
\[
\rho_{B|\Pi_i^b} = \text{Tr}_A\left(\frac{(\Pi_i^b \otimes \mathbb{I})\rho_{AB}(\Pi_i^b \otimes \mathbb{I})}{p_{\Pi_i^b}}\right).
\]
The $l_1$-norm of coherence in the $\sigma_i$ basis is calculated as
\[
C_{l_1}^{\sigma_i}(\rho) = \sum_{L \neq R} |\langle L | \rho | R \rangle|,
\]
where $\{|L\rangle, |R\rangle\}$ are the eigenstates of the Pauli operator $\sigma_i$.
The criterion in Eq.~\eqref{23} determines whether the bipartite state possesses non-local advantage of quantum coherence,
which represents a stronger quantum correlation than entanglement.

\subsection{Bell Nonlocality}
BN is the fundamental manifestation of the violation of local realism in quantum mechanics,
and the Clauser-Horne-Shimony-Holt (CHSH) inequality is the most widely used criterion for detecting BN in two-qubit systems.
For a bipartite quantum state $\rho_{AB}$, the CHSH inequality is given by \cite{shadbolt2012guaranteed}
\begin{equation}\label{24}
B(\rho_{AB}) = |\langle B_{\text{CHSH}} \rangle| = |\text{Tr}(\rho_{AB} B_{\text{CHSH}})| \leq 2,
\end{equation}
where the CHSH Bell operator is defined as
\begin{equation}\label{25}
B_{\text{CHSH}} = \vec{a} \cdot \vec{\sigma} \otimes (\vec{b} + \vec{b}')\cdot\vec{\sigma} + \vec{a}' \cdot \vec{\sigma} \otimes (\vec{b} - \vec{b}')\cdot\vec{\sigma},
\end{equation}
with the following parameters:
\begin{itemize}
\item[(1)] $\vec{a}, \vec{a}'$ and $\vec{b}, \vec{b}'$ are unit vectors representing the measurement directions of the two observers;
\item[(2)] $\vec{\sigma} = (\sigma_x, \sigma_y, \sigma_z)$ denotes the set of Pauli matrices;
\item[(3)] $\otimes$ represents the tensor product for independent measurements on each subsystem.
\end{itemize}

Violation of the CHSH inequality ($B>2$) indicates that the quantum state exhibits Bell nonlocality.
For any two-qubit state, the maximum violation value is determined by the correlation tensor and can be computed as
\begin{equation}\label{26}
B_{\text{max}}(\rho_{AB}) = 2\sqrt{\max_{i<j}(u_i + u_j)},
\end{equation}
where $u_i$ ($i=x,y,z$) are the eigenvalues of the matrix $T^\dagger T$, and the correlation tensor $T$ is given by
\[
T_{ij} = \text{Tr}[\rho_{AB} (\sigma_i \otimes \sigma_j)].
\]
The general form of a two-qubit density matrix in the Bloch representation is
\[
\rho_{AB} = \frac{1}{4}\left(\mathbb{I} \otimes \mathbb{I} + \vec{a} \cdot \vec{\sigma} \otimes \mathbb{I} + \mathbb{I} \otimes \vec{b} \cdot \vec{\sigma} + \sum_{i,j=1}^3 T_{ij} \sigma_i \otimes \sigma_j\right).
\]

Notably, Bell nonlocality is symmetric and bidirectional: exchanging the roles of the two subsystems does not change the value of $B_{\text{max}}$,
which is an important property for analyzing quantum correlations in curved spacetime.
\section{Bipartite Fermionic Entanglement and Nonlocal Correlations in EGB Black Hole Spacetime}
\label{sec:4}
Following the setup established in previous sections, we consider the physical scenario in which Rob moves toward the black hole and remains stationary outside the event horizon via weak acceleration, while Alice freely falls along a geodesic and crosses the horizon within a finite proper time. Initially, the two observers share a maximally entangled Bell state defined in the Minkowski spacetime far from the gravitational background:
\begin{equation}\label{27}
|\phi \rangle_{M} = \frac{1}{\sqrt{2}}\left( |0\rangle_{M}^{\mathrm{A}}|0\rangle_{M}^{\mathrm{R}} + |1\rangle_{M}^{\mathrm{A}}|1\rangle_{M}^{\mathrm{R}} \right),
\end{equation}
where the first qubit corresponds to Alice’s  mode and the second to Rob’s mode, with the subscript $M$ denoting Minkowski modes. Owing to the Unruh–Hawking effect induced by curvature and acceleration, the initial perfect entanglement between the two subsystems is degraded from Rob’s perspective. We assume that both cavities are initially empty and support two orthogonal single-frequency modes within the single-mode approximation; each mode is excited to a Fock state at the coincidence point of the two observers.

Since Rob is confined to the accelerated frame and described by Schwarzschild coordinates, his field modes must be decomposed according to the Bogoliubov transformation in Eq.~\eqref{22}. Expanding the bipartite Bell state with Minkowski modes for Alice and Schwarzschild modes for Rob yields a tripartite density matrix $\rho_{A,\mathrm{I},\mathrm{II}}$. As Region II is causally inaccessible to Rob, we perform a partial trace over the unobservable degrees of freedom in Region II, resulting in the reduced mixed bipartite density matrix $\rho_{A,\mathrm{I}}$.

We further construct the Bell state from the eigenmodes of the fermionic Dirac field propagating in the EGB gravitational background. Starting from the asymptotic flat region where entanglement is initially shared, Alice proceeds along a free-fall trajectory into the black hole, whereas Rob maintains a stationary position outside the horizon. Using the mode transformation relation in Eq.~\eqref{21}, we expand Rob’s local modes in the Schwarzschild basis and derive the full tripartite density operator:
\begin{equation}\label{28}
\begin{aligned}
\rho_{\mathrm{A,I,II}} = & \frac{1}{2} \big( \cos^2 \zeta \, |000\rangle \langle 000| + \sin^2 \zeta \, |011\rangle \langle 011| + |110\rangle \langle 110| \big) \\
& + \frac{1}{2} \big( \cos \zeta \sin \zeta \, |000\rangle \langle 011| + \cos \zeta \, |000\rangle \langle 110| + \sin \zeta \, |011\rangle \langle 110| + \mathrm{H.C.} \big).
\end{aligned}
\end{equation}
After tracing out the causally disconnected Region II, the physically accessible bipartite density matrix reads
\begin{equation}\label{29}
\rho_{A ,\mathrm{I}} = \frac{1}{2}
\begin{pmatrix}
\cos^2\zeta & 0 & 0 & \cos \zeta \\
0 & \sin^2\zeta & 0 & 0 \\
0 & 0 & 0 & 0 \\
\cos \zeta & 0 & 0 & 1
\end{pmatrix}.
\end{equation}

 {Even though the algebraic form of the reduced density matrix is analogous to that of Schwarzschild spacetime, the Hawking temperature $T$ here contains extra free parameters $\alpha$ and $d$, which do not exist in Einstein gravity. Such parameter-dependent thermal input leads to distinct parametric evolutions of NAQC and BN that cannot be reproduced by conventional black hole models, which endows our calculations with independent physical value. Mathematically, the bipartite density matrix in Eq. \eqref{29} shares the same structure as the result derived for 4-dimensional Schwarzschild black holes. This universal matrix form comes from the two-mode fermionic Bogoliubov transformation under single-mode approximation, which only relies on the thermal parameter $\zeta$. However, $\zeta$ is determined by $T(\alpha,d,r_{h})$ in EGB spacetime, carrying high-dimensional and high-curvature information unavailable in conventional Schwarzschild geometry.}

Substituting the reduced density matrix into the criteria introduced in Eqs.~\eqref{23}--\eqref{26}, we obtain the analytical expressions for NAQC and Bell nonlocality:
\begin{equation}\label{30}
N^{l_{1}}(\rho_{A,\mathrm{I}})=\frac{1}{2}+|\cos \zeta|+\frac{1}{2}|\cos 2\zeta|+\sqrt{\sin^4\zeta+\cos^{2}\zeta},
\end{equation}
\begin{equation}\label{31}
B_{\mathrm{max}}(\rho_{A,\mathrm{I}})=2\sqrt{2}\,|\cos \zeta|.
\end{equation}

\begin{figure}[htbp]
\centering
\includegraphics[width=1\textwidth]{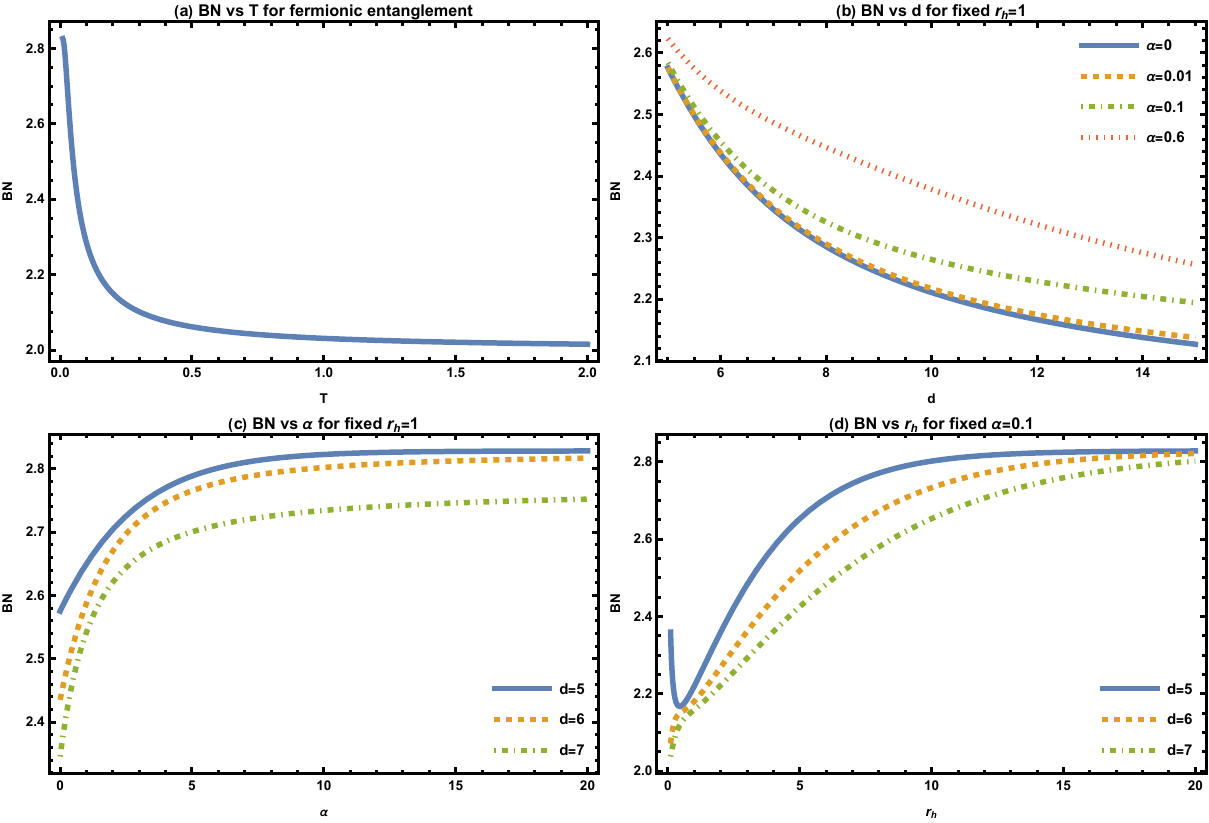}
\caption{$B_{\mathrm{max}}(\rho_{A,\mathrm{I}})$ characterizing Bell nonlocality for fermionic entanglement.}
\label{fig:BN}
\end{figure}

Figure~\ref{fig:BN} illustrates the evolution of Bell nonlocality for fermionic entanglement under varying physical parameters. Throughout all numerical calculations, we fix $\omega=0.01$, a low-energy value that preserves the validity of the single-mode approximation and does not alter the qualitative behavior of the correlations. Figure~\ref{fig:BN}(a) presents $B_{\mathrm{max}}$ as a function of the Hawking temperature $T$. In the zero-temperature limit, the system remains maximally entangled with $B_{\mathrm{max}}=2\sqrt{2}$. As $T$ increases, thermal radiation gradually suppresses Bell nonlocality; nevertheless, $B_{\mathrm{max}}$ always exceeds $2$, indicating persistent violation of the CHSH inequality and the survival of Bell nonlocality at arbitrary finite temperature. Figure~\ref{fig:BN}(b) displays the dependence of BN on spacetime dimension $d$ for fixed $\alpha$ and $r_h=1$. Bell nonlocality decreases monotonically with increasing dimension, while a larger Gauss--Bonnet coupling $\alpha$ enhances nonlocality and weakens the decaying trend. This behavior originates from the antigravitational correction of the positive Gauss--Bonnet term, which suppresses curvature-induced decoherence. Figure~\ref{fig:BN}(c) shows that BN increases monotonically with $\alpha$ and asymptotically saturates at a constant value; meanwhile, higher dimensionality consistently reduces the magnitude of nonlocality, consistent with the trend in Fig.~\ref{fig:BN}(b). Figure~\ref{fig:BN}(d) describes the variation of BN with horizon radius $r_h$ for fixed $d$ and $\alpha=0.1$. For $d>5$, BN grows monotonically with $r_h$ and converges to a constant; for $d=5$, BN exhibits a slight initial drop at extremely small $r_h$ before increasing. This distinction can be interpreted from the different temperature expressions in Eqs.~\eqref{10} and~\eqref{11}, where $T$ diverges at $r_h\to0$ for general $d$, whereas $T\to0$ in the five-dimensional limit.

\begin{figure}[htbp]
\centering
\includegraphics[width=1\textwidth]{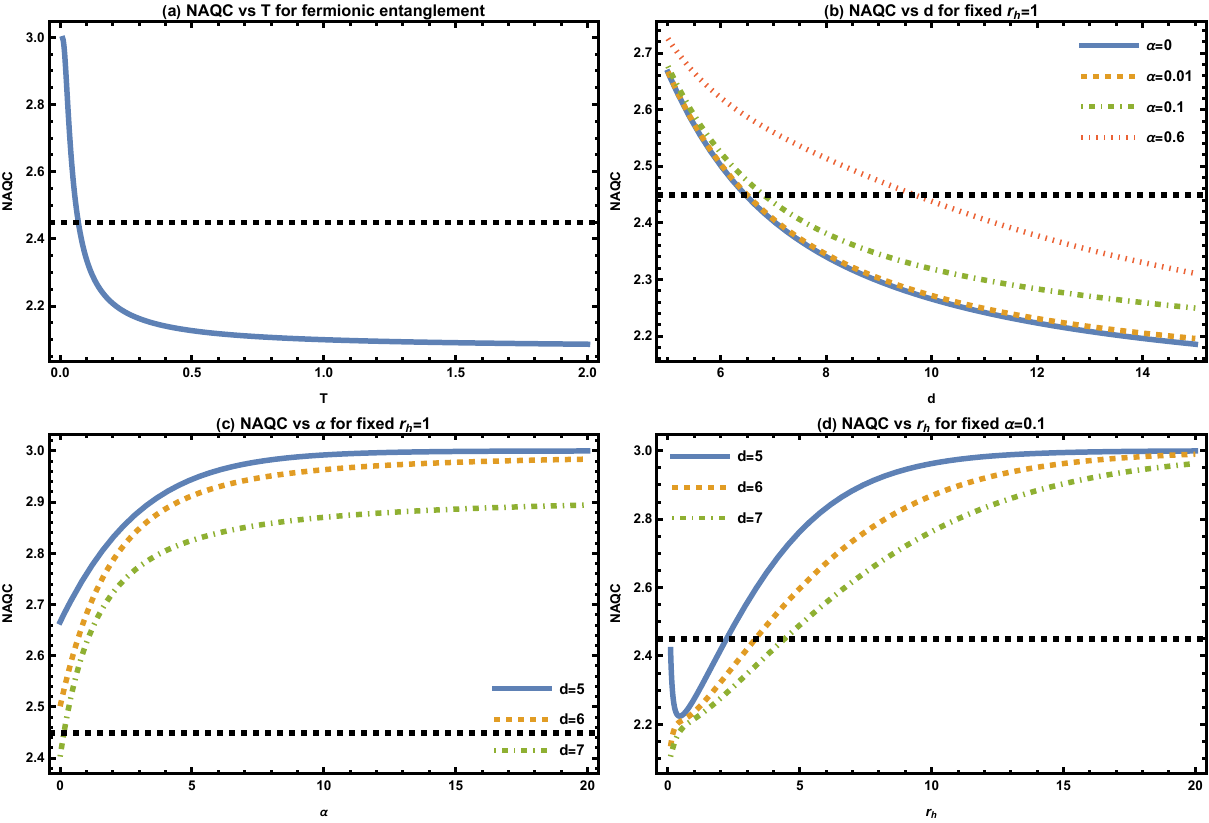}
\caption{$N^{l_{1}}(\rho_{A,\mathrm{I}})$ characterizing the NAQC for fermionic entanglement (from Alice to Rob).}
\label{fig:ZNAQC}
\end{figure}

Figure~\ref{fig:ZNAQC} demonstrates the corresponding behavior of NAQC. Overall, the trends closely resemble those of Bell nonlocality, with a crucial difference: NAQC vanishes once $N^{l_{1}}(\rho_{A,\mathrm{I}})\leq\sqrt{6}$, defining a strict threshold for nonlocal coherence advantage. Figure~\ref{fig:ZNAQC}(a) shows that NAQC decays monotonically with rising Hawking temperature and disappears beyond a critical temperature. Figure~\ref{fig:ZNAQC}(b) reveals that NAQC is suppressed with increasing spacetime dimension and ultimately vanishes; a larger $\alpha$ extends the critical dimension at which NAQC disappears. Figure~\ref{fig:ZNAQC}(c) confirms that NAQC grows with $\alpha$ and saturates asymptotically, while higher dimensionality diminishes the achievable coherence advantage.  {The magnitude $|\cos{\zeta}|$ carries clear physical meaning: it quantifies the proportion of original maximal entanglement that survives Hawking thermal decoherence, while $\sin\zeta$ represents the thermal particle excitation component induced by the Unruh-Hawking effect.A larger $|\cos\zeta|$ corresponds to weaker thermal mixing and less entanglement degradation. Since $B_{max}=2\sqrt2|\cos\zeta|$, Bell nonlocality is linearly proportional to $|\cos\zeta|$ and rises monotonically with it. All terms constructing the NAQC indicator also increase with $|\cos\zeta|$, so stronger nonlocal coherence advantage emerges as $|\cos\zeta|$ grows. When the Gauss-Bonnet coupling constant $\alpha$ rises, stronger antigravitational higher-curvature corrections weaken the local spacetime curvature near the event horizon and reduce the black hole’s surface gravity; following the geometric definition of Hawking temperature $T=\frac{1}{4\pi}\frac{df}{dr}\big|_{r_h}$, smaller surface gravity yields a lower Hawking temperature $T$, and the relation $\tan\zeta=e^{-\pi\omega/T}$ consequently increases $\zeta$ and elevates $|\cos\zeta|$, where $|\cos\zeta|$ quantifies the fraction of original entanglement preserved against Unruh-Hawking thermal decoherence, and since both Bell nonlocality and NAQC monotonically grow with $|\cos\zeta|$, larger $\alpha$ ultimately enhances all measured nonlocal quantum correlations by suppressing curvature-induced thermal decoherence. Meanwhile, higher spacetime dimension $d$ amplifies horizon curvature and lifts $T$, creating a competitive relationship between dimensionality and GB correction that cannot be observed in 4-dimensional general relativity.} Figure~\ref{fig:ZNAQC}(d) indicates that NAQC is absent for small horizon radii and emerges gradually as $r_h$ increases, finally approaching a stable limit; larger dimensions delay the emergence of detectable NAQC.

We further examine the directional asymmetry of these correlations by swapping the roles of Alice and Region I. Bell nonlocality remains invariant under exchange, reflecting its fundamental symmetry, whereas NAQC exhibits directional dependence. The corresponding reversed density matrix is
\begin{equation}\label{32}
\rho_{\mathrm{I},A} = \frac{1}{2}
\begin{pmatrix}
\cos^2\zeta & 0 & 0 & \cos \zeta \\
0 & 0 & 0 & 0 \\
0 & 0 & \sin^2\zeta & 0 \\
\cos \zeta & 0 & 0 & 1
\end{pmatrix},
\end{equation}
and the associated NAQC reads
\begin{equation}\label{33}
N^{l_{1}}(\rho_{\mathrm{I},A})=2|\cos \zeta|+\cos^{2}\zeta.
\end{equation}

\begin{figure}[htbp]
\centering
\includegraphics[width=1\textwidth]{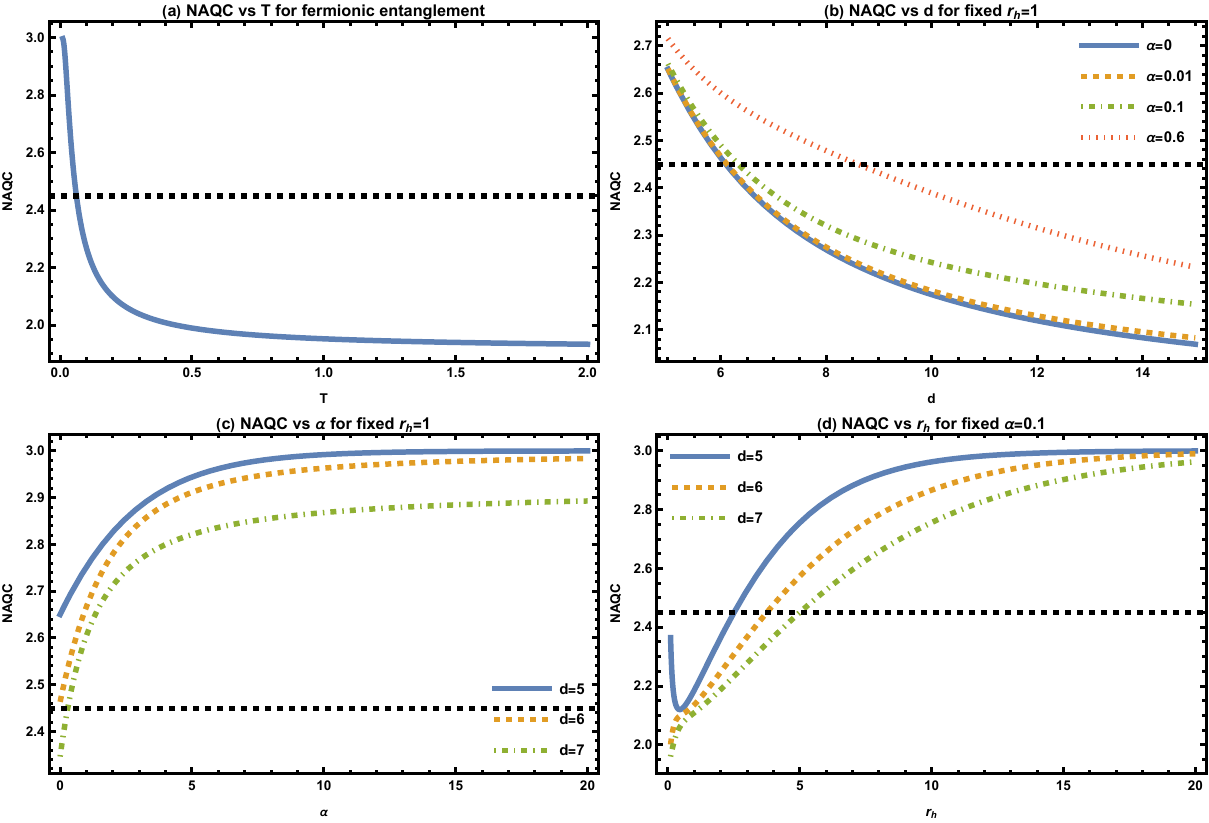}
\caption{$N^{l_{1}}(\rho_{\mathrm{I},A})$ characterizing the reversed NAQC (from Rob to Alice).}
\label{fig:FNAQC}
\end{figure}

As plotted in Fig.~\ref{fig:FNAQC}, the reversed NAQC shares qualitatively similar parameter dependence with the original NAQC but differs quantitatively. Both NAQC and Bell nonlocality are weakened monotonically by increasing Hawking temperature, unambiguously demonstrating that thermal fluctuations from the Hawking effect degrade quantum nonlocal resources in curved higher-curvature spacetimes. Moreover, the fact that NAQC vanishes at finite temperature while Bell nonlocality persists rigorously verifies the hierarchical relation: NAQC constitutes a stronger nonlocal quantum resource compared with standard Bell nonlocality, and this hierarchy survives in the EGB gravitational background.

 {When $\alpha\rightarrow0$, the EGB gravity reduces to higher-dimensional Schwarzschild gravity, and our temperature expression Eq. \eqref{10} recovers the standard Hawking temperature for $d$-dimensional Schwarzschild black holes. All curves converge to known results reported in previous literature, which confirms the self-consistency of our derivations.}

\section{Conclusion}
\label{sec:5}
In this work, we systematically investigate bipartite nonlocal correlations of non-interacting fermionic fields in the spacetime of a $d$-dimensional spherically symmetric Einstein-Gauss-Bonnet  black hole. We consider a physically motivated scenario involving two observers: Alice, who freely falls into the black hole and thus occupies an inertial frame described by Kruskal coordinates, and Rob, who maintains a stationary position outside the event horizon via acceleration and resides in a non-inertial frame described by Schwarzschild-like coordinates. By extending the analysis of Schwarzschild black holes to the higher-curvature EGB framework, we derive the Bogoliubov transformation that connects the field modes in the inertial (Kruskal) and non-inertial (Schwarzschild) frames, accounting for the spacetime curvature and acceleration-induced Unruh–Hawking effect.

Starting from a maximally entangled Bell state shared by the two observers in the asymptotic flat region (far from the black hole), we construct the quantum state using single-mode Fock states of the fermionic Dirac field and examine two key quantifiers of non-local correlations: the Nonlocal Advantage of Quantum Coherence  and Bell Nonlocality. Our analytical and numerical results reveal that the hierarchical relationship between NAQC and BN—originally established in flat spacetime—persists in the curved EGB background: NAQC serves as a stricter criterion for non-local quantum resources than BN, confirming that NAQC constitutes a subset of BN even under the influence of gravitational thermal effects and spacetime curvature.

A central finding of our study is that both NAQC and BN degrade monotonically with increasing Hawking temperature, which directly demonstrates that thermal radiation generated by the Hawking effect weakens non-local quantum correlations in curved higher-curvature spacetimes. We further explore the dependence of these non-local correlations on key EGB black hole parameters, including the spacetime dimension $d$, Gauss-Bonnet coupling constant $\alpha$, and horizon radius $r_h$. Our numerical simulations  show that NAQC and BN exhibit analogous variation patterns with these parameters: both are suppressed by higher spacetime dimensionality, enhanced by a larger positive $\alpha$ (attributed to the anti-gravitational effect of the Gauss-Bonnet term that mitigates decoherence), and modulated by the horizon radius in a dimension-dependent manner. Additionally, we confirm the directional asymmetry of NAQC—its magnitude changes when the roles of Alice and Rob are swapped—while BN remains symmetric, consistent with its fundamental property of bidirectional invariance.  {Beyond the universal monotonic decay of quantum correlations with growing Hawking temperature widely reported in previous literature, we identify novel high-curvature and dimensional effects exclusive to EGB spacetime, including dimension-split $r_{h}$ behavior and decoherence suppression from positive GB coupling, which significantly extends the understanding of relativistic quantum resources beyond standard GR backgrounds. The $\alpha\rightarrow0$ limit of our numerical results perfectly reproduces established results for higher-dimensional Schwarzschild black holes, validating the whole theoretical framework.}

These results not only extend the understanding of quantum resource theory to strong gravitational fields with higher-curvature corrections but also provide new insights into the interplay between quantum nonlocality and gravitational physics. By verifying the robustness of the NAQC-BN hierarchy in the EGB black hole spacetime, our work contributes to the ongoing effort to reconcile quantum mechanics and general relativity, shedding light on the quantum nature of black holes and the behavior of quantum resources in extreme gravitational environments. Future work could extend this analysis to interacting fermionic fields or explore non-local correlations in other higher-curvature gravity models.

\section*{Acknowledgements}
This work benefits from the high performance computing clusters at School of Physics and Optoelectronic Engineering, Yangtze University.

\bibliographystyle{unsrt} 
\bibliography{references.bib}

\end{document}